\documentclass[runningheads]{llncs}
\usepackage{graphicx}

\usepackage{color,soul,booktabs,eurosym, hyperref}

\begin{document}
    \title{Keep on Running!\\An Analysis of Running Tracking Application Features and their Potential Impact on Recreational Runner's Intrinsic Motivation}
\titlerunning{Keep on Running! Running Tracking Applications and Intrinsic Motivation}
%

%
%

\author{Dorothea Gute \and
Stephan Schl\"{o}gl\orcidID{0000-0001-7469-4381} \and
Aleksander Groth} 
\authorrunning{Gute et al.}
%
\institute{MCI -- The Entrepreneurial University\\Innsbruck, Austria\\
Dept. Management, Communication \& IT \\
\email{stephan.schloegl@mci.edu}\\
\url{https://www.mci.edu}} 

%
\maketitle              
\begin{abstract}
Physical activity is known to help improve and maintain one's health. In particular, recreational running has become increasingly popular in recent years. Yet, lack of motivation often interferes with people's routines and thus may prohibit regular uptake. This is where running tracking applications are frequently used to overcome one's weaker self and offer support. While technology artifacts, such as sport watches or running applications, usually count as extrinsic drivers, they can also impact one's intrinsic motivation levels. The aim of this study was thus to investigate upon the motivational impact of distinct features found within applications specifically used for running. Focusing on the 22 most famous running applications, a semi-structured, problem-centered interview study with $n=15$ recreational runners showed that intrinsic motivation is stimulated from diverting runners, aiding them in their goal setting, decreasing their efforts, improving and sharing their run performance, allowing them to receive acknowledgements, as well as providing them with guidance, information, and an overall variety in their training routines. 



%
\end{abstract}
\section{Introduction}\label{sec:intro}
Physical activity has various benefits on humans' health which is continuously outlined by the World Health Organization (WHO)\footnote{Online: \url{https://www.who.int/news-room/fact-sheets/detail/physical-activity} [accessed: October 10\textsuperscript{th}, 2021]} as well as health professionals all around the world. Backed by a consequently increased health awareness, running has become a particularly popular recreational activity in our modern society. It requires little expertise and may be exercised in a variety of different environments~\cite{scheerder2015running}. Yet, while regular engagement in running activities may boost one’s overall health status, lack of time, motivation or enthusiasm are often cited as significant adversaries~\cite{burgess2017determinants}. Recreational runners in particular have to bring in and sustain a high level of intrinsic motivation in order to keep them going~\cite{wijnalda2005personalized}.

Motivated people seem to be energized, while a lack of motivation is often described as `missing the drive' or lacking inspiration to carry out an action~\cite{ryan2000intrinsic}. Moreover, research has shown that intrinsic as opposed to extrinsic motivation leads to \textit{``greater interest, greater effort, better performance, a more positive emotional tone, higher instances of flow, higher self-esteem, better adjustment, [and] greater satisfaction''}~\cite[p. 6]{pelletier2001associations}. This can be explained by the observation that intrinsically motivated individuals take action because they find it enjoyable and interesting, whereas extrinsically motivated people do so because of an outcome said action will have -- one, which is usually not connected to the action itself~\cite{ryan2000intrinsic}. With respect to practicing sports, intrinsically motivated people exercise because they derive pleasure and amusement from this activity, while extrinsically motivated athletes rather engage in sports to receive tangible advantages such as objects (e.g., prices), social acknowledgment, or to prevent punishment~\cite{vallerand1999integrative}.

Applications for tracking runs, usually connected to different types of wearables (e.g., smartwatches), are widely used to help fight runners' lack of motivation and consequently, to a certain extend, aim to trigger behavioural change in people. 
Yet, in order for these applications to have a lasting effect they \textit{``need to create enduring new habits, turning external motivations into internal ones''}~\cite[p. 1]{patel2015wearable}.
Inspired by this endeavor, the aim of this work was to investigate, how well the most popular of these running applications live up to this challenge of affecting a runners' motivation. In other words, we were investigating: 
\vspace{0.2cm}
\begin{center}
\textit{Which running tracking application features impact on recreational runners’ intrinsic motivation, and how?}
\end{center}
\vspace{0.2cm}
Our report starts with a discussion of the relevant theoretical framework for the respective investigation in Section~\ref{sec:relwork}. Then, Section~\ref{sec:methodology} describes our methodological approach including our sampling strategy. Section~\ref{sec:results} summarizes and discusses the gained results, before Section~\ref{sec:conclusion} concludes and provides some directions for further research.

\section{Theoretical Framework and Related Work}\label{sec:relwork}
Intrinsic motivation is critical for sustaining physical activity due to it being perceived as enjoyable~\cite{richard1997intrinsic}. Further, it has been found that being extrinsically motivated, e.g. to win an award or a competition, decreases one’s intrinsic motivation, for one perceives the action as controlling rather than enjoyable~\cite{deci1981trying,deci1999meta}. Moreover, when individuals engage in an activity which has been extrinsically motivated, they endure it less than when this external motivator is removed~\cite{pelletier2001associations}. However, when these extrinsic rewards are not dependent on the activity or a specific achievement, their influence and consequently their controlling effect decreases. 
It may thus be argued, that intrinsic and extrinsic motivation are not additive~\cite{deci2008facilitating}.

\subsection{Self Determination Theory}
Deci \& Ryan~\cite{deci2013intrinsic} suggest that intrinsic motivation becomes effective when an activity is experienced as being self-determined, and at the same time improbable or inhibited, when said activity is experienced as being controlled externally. To this end, Self Determination Theory (SDT) assumes that humans are inherently active, motivated, knowledge desiring and keen on being successful. One's social environments can endorse a person's self-determined nature, but at the same time also impede it. According to SDT, humans generally strive to feel \emph{competent}, \emph{autonomous}, and \emph{related to others} and that social contexts can influence one’s fulfillment of these needs positively as well as negatively~\cite{deci2008facilitating,ryan2000self}.
While competence and autonomy have the strongest influence on intrinsic motivation, relatedness is less influential, yet still supportive (ibid.).

Vallerand distinguishes between three types of intrinsic motivation, where \textit{``intrinsic motivation to know can be defined as engaging in an activity for the pleasure and satisfaction one experiences while learning, exploring, or trying to understand something new}~\cite[p. 280]{vallerand1997toward}. In the context of sports, an example for this type of motivation would be a basketball player who enjoys learning new offensive moves~\cite{Vallerand2012IntrinsicAE}. Deriving pleasure from exceeding one’s own accomplishments or creating something, on the other hand, is considered \textit{``intrinsic motivation to accomplish things''}. Lastly, taking action to experience enjoyable sensations is known as \textit{``intrinsic motivation to experience stimulation''} (ibid.). 

\subsection{Hierarchical Model of Intrinsic Motivation}
Enhancing SDT, the Hierarchical Model of Intrinsic Motivation (HMIM) distinguishes three hierarchical levels of motivation, i.e., the \emph{global}, \emph{contextual}, and \emph{situational} level. Motivation at the global level understands humans to generally act according to one inherent form of motivation, disregarding the context or situation they find themselves in~\cite{vallerand1997toward}. For example, an athlete who is intrinsically motivated towards sports behaves the same way towards their education, work, friends and leisure time~\cite{schuler2020intrinsische}. On the other hand, adapting one's motivation towards a certain context is considered being motivated at the context level~\cite{vallerand1997toward}. That is, a person being amotivated (i.e., not motivated) towards sports can simultaneously be intrinsically motivated towards their occupation~\cite{schuler2020intrinsische}. Finally, situational motivation is experienced while undertaking an action. 
For instance, an athlete can be intrinsically motivated to study a difficult type of movement, so as to fulfill an individual need for competence~\cite{schuler2020intrinsische}.



Furthermore, perceived \emph{competence}, \emph{autonomy} and \emph{relatedness} mediates the influence of social factors on these three levels. 
A stable social network (e.g., one’s parents) that supports one's need for autonomy, may be considered a global factor, whereas the support of a trainer may be considered a contextual factor, and a positive performance review for a distinct training unit a situational factor~\cite{schuler2020intrinsische}.



Another rule of HMIM assumes that hierarchically higher forms of motivation influence the lower levels, which signifies a so-called \emph{top-down effects}. Consequently, contextual motivation impacts situational motivation stronger than global motivation does and global motivation influences situational motivation~\cite{vallerand1997toward}. In other words, a person who has high levels of intrinsic motivation on a global level will most likely possess the same type of motivation in several contexts, e.g., sports, and accordingly in specific sport situations, e.g., a particular exercise during training~\cite{schuler2020intrinsische}.

On the other hand, motivational changes are taken into consideration by suggesting that motivation at one level can have a \emph{bottom-up effect} at the next higher level (ibid.). That is, intrinsic motivation at a contextual level (i.e., motivation for doing sport) can be increased during an athlete's individual training by providing him/her with a scope of action, performance feedback and a pleasant atmosphere (situational level). 

\begin{figure}
    \centering
    \includegraphics[width=\textwidth]{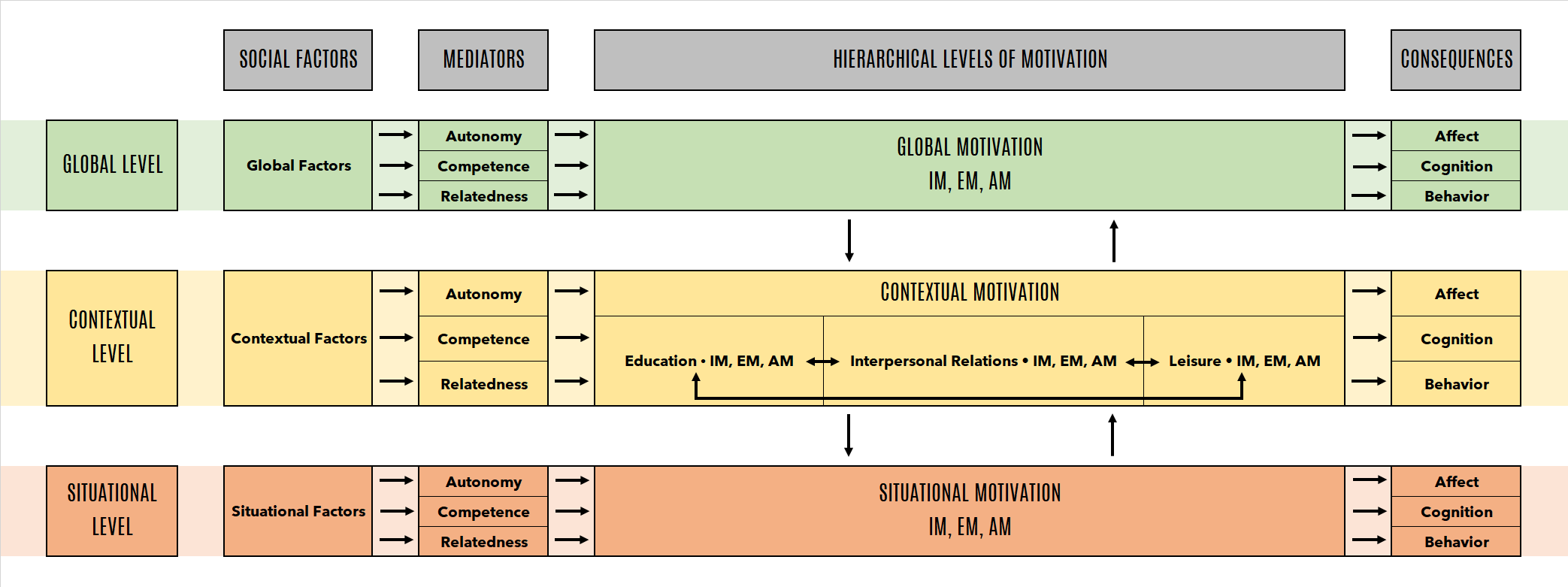}
    \caption{Hierarchical Model of Intrinsic Motivation adapted from Vallerand \& Lalande~\cite[p. 46]{vallerand2011mpic}}
    \label{fig:hmim}
\end{figure}

Concludingly, HMIM assumes that motivation causes important consequences concerning \emph{affect}, \emph{cognition}, and \emph{behavior}, leading to higher levels of creativity and learning, increased interest, positive emotions, and greater task persistence~\cite{vallerand1997toward}. Fig.~\ref{fig:hmim} summarizes the different hierarchy levels of HMIM and their interrelations.  


\subsection{Goal Setting}
Individual goals guide actions, particularly in athletics~\cite{elbe2020motivation}.
Difficult and specific goals are given more attention, exertion, and perseverance and therefor have a higher probability of being achieved~\cite{locke1990theory}. 
As a result, a sense of commitment and determination to overcome or plan around potential obstacles~\cite{elbe2020motivation} is created.
Achievement Goal Theory (AGT) defines achievement behavior as \textit{``behavior in which the goal is to develop or demonstrate – to self or to others – high ability, or to avoid demonstrating low ability''}~\cite[p. 328]{nicholls1984achievement}. Respective goals may be \emph{ego-oriented} or \emph{task-oriented}, where the former focuses on the performance results (e.g., in the sense of a competition or social comparison) and the latter on learning and task completion (ibid.). 
Performance-oriented people aim to demonstrate their skills to others by competing with them, which may result in disappointment, frustration, and loss of motivation, and generally have a negative effect on one’s perceived competence~\cite{duda2001achievement,elbe2020motivation}.
Task-oriented people, on the other hand, orient towards individual standards, and are thus less afraid of failure, which leads to a higher level of perceived competence. 
In a sports context, task-orientation is further associated with increased joy, intrinsic interest, and willingness for endeavor. 
Transferred to motivation, this means that focusing on one’s tasks, i.e., the activity itself, rather than on comparing oneself to others, fosters intrinsic motivation~\cite{elbe2020motivation}. 
And although McAuley \& Tammen found that winning can increase intrinsic motivation, one’s perceived level of competence contributes even stronger~\cite{mcauley1989effects}. 

Furthermore, Ryan \& Deci (2000b) state that self-reporting one’s achievements increases intrinsic motivation~\cite{ryan2000self} and receiving feedback contributes to one’s feeling of competence~\cite{suh2015effects}; although when said feedback is perceived as pressuring or controlling, it may undermine intrinsic motivation~\cite{smith1975social}.

\subsection{Flow}
The concept of \emph{flow} describes an optimal state of motivation, representing a prototype of intrinsic motivation~\cite{schuler2020intrinsische}. Csikszentmihalyi aimed to understand an optimal experience and found that a flow effect is achieved as \textit{``a state of concentration so focused that it amounts to absolute absorption in an activity''}~\cite[p. 1]{csikszentmihalyi1990flow}. In a sports context, flow states occur when athletes are neither concerned about rational reasons against an activity, e.g., time that must be invested, nor about reasons in favor of engaging in that activity, like increased health~\cite{schuler2020intrinsische}. Participation is rather encouraged by the activity itself and in doing so leading to a feeling of enjoyment~\cite{csikszentmihalyi1990flow}.
Flow requires \emph{a clear set of goals} and a balance between perceived challenges and perceived skills, as well as clear and immediate feedback; goals help direct one’s behavior by conveying attention. The focus lies on the balance of challenge and skill requirements, for when the former are too demanding, one becomes too concerned about failure. When challenges are not perceived at all, one loses interest. Immediate feedback advises on ones' progress and possible adjustment of actions resulting in empowerment~\cite{csikszentmihalyi2005handbook}.

The four most important characteristics of flow include: (1) concentration on the task at hand, (2) merging of action and awareness, (3) having a sense of control over the activity, and (4) transformation of time~\cite{schuler2020intrinsische}. Csikszentmihalyi argues that, when experiencing flow, one concentrates strongly on the given task and is able to suppress unpleasant facets of their life. The action itself merges with one’s awareness and functions almost automatically. Being in control during the activity, means precisely exercising control and, therefore, providing a decrease in stress during demanding situations. Also, time seems to pass differently from how it normally does when experiencing flow. Most commonly, sequences of the action are perceived as if time would stand still, however, the experience itself is perceived as if time was passing remarkably fast~\cite{csikszentmihalyi1990flow}. While flow is known to enable peak performances, research has shown that it can also have negative effects, such as risk taking or underestimating. 

\subsection{Running Applications}
Smartphone-based applications that track physical activities such as running, usually consist of four components~\cite{ahtinen2008tracking}:
\begin{enumerate}
    \item a logger which measures and stores exercise related data,
    \item a virtual personal trainer that adds an analysis functionality and provides visual and audible performance feedback,
    \item gaming and entertainment functions, and
    \item community and social sharing features.
\end{enumerate}
They allow for individual use (e.g., goal setting) and support social interaction (e.g., sharing and feedback). Thus, these applications aim to induce both intrinsic as well as extrinsic motivation by either engaging users in taking action \textit{``due to feelings of pleasantness and satisfaction inherent in the activity [...] [or by receiving] recognition or approval from others (or the app itself)''}~\cite[p. 1428]{hosseinpour2019your}. Investigating the potential for behavioural change based on the SDT, Villalobos-Zúñiga \& Cherubini~\cite{VillalobosZiga2020} found that certain application features, depending on their perceived level of external control, can either positively or negatively influence intrinsic motivation (e.g., reminders, activity feedback, motivational messages, rewards, etc.). With a particular focus on running, Bauer \& Kriglstein~\cite{bauer2015analysis} further suggest that the music and audio feedback, different types of visualization, as well as an app's competition and comparison feature helps foster athlete's motivation. 

Our goal was to investigate whether those suggestions are still valid and how the different features are perceived by a specific target group; i.e., recreational runners.

\section{Methodology}\label{sec:methodology}

Building upon the work of Bauer \& Kriglstein~\cite{bauer2015analysis}, we focused on the 22 top running tracking applications currently available for iOS and Android OS, and analyzed their features via $n=15$ guided, semi-structured, problem-centered interviews. Our sampling approach used a screening questionnaire to identify participants with low intrinsic motivation towards running. We then used Kuckartz~\cite{kuckartz2014qualitative} structuring content analysis approach and a slight adaptation of Gutman's laddering technique~\cite{gutman1982means} to explore interviewees’ inherent attitudes as to how various application features may impact on their intrinsic motivation to run. 

\subsection{App Selection}
\begin{figure}
    \centering
    \includegraphics{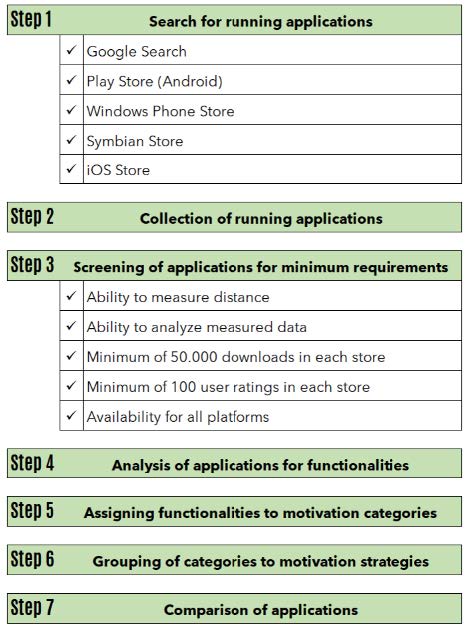}
    \caption{Analysis of Motivation Strategies in Running Tracking Applications adapted from Bauer \& Kriglstein~\cite[p. 2]{bauer2015analysis}}
    \label{fig:apps}
\end{figure}

Bauer \& Kriglstein~\cite{bauer2015analysis} analyze \textit{``the top free running tracking applications on the market with their functionalities to identify which motivation strategies are supported by these applications in 2013''}~[p. 2]. They chose to include three feature elements which they thought would motivate runners, i.e., \emph{music and audio feedback}, \emph{visualization}, as well as \emph{competition and comparison with others}. As Fig.~\ref{fig:apps} shows, the authors’ analysis consisted of seven steps. 

First, a search for running applications on Google and in the different app stores (i.e., Play Store, Windows Phone Store, Symbian Store, iOS Store) had been conducted. 
After collecting these applications, they screened them for five minimum requirements (i.e., ability to measure distance, ability to analyze measured data, a minimum of \emph{50K} downloads in each store, a minimum of \emph{100} users ratings in each store, and availability for all platforms). Subsequently, the apps' functionalities were tested, assigned to different motivation categories, grouped into respective strategies and eventually compared.

Our approach followed the same example, yet as 8 years have past since Bauer \& Kriglstein's work, we integrated some additonal limitations.
That is, based on statistics by Appfigures \& VentureBeat\footnote{Online: \url{https://www.statista.com/statistics/276623/number-of-apps-available-in-leading-app-stores/} [accessed: January 13\textsuperscript{th}, 2021]}, Google's Play store was the app store with the most available apps in the first quarter of 2021 (3.482.452 applications), followed by Apple's iOS store (2.226.823 applications). Consequently, we decided to focus on these two app stores in combination with a Google search for \textit{``Free running tracking applications''}, and to exclude the Windows phone store due to its minor size (only 669.000 available apps)\footnote{Online: \url{https://www.statista.com/statistics/276623/number-of-apps-available-in-leading-app-stores/} [accessed: January 13\textsuperscript{th}, 2021]} as well as the Symbian store, as Nokia terminated its development and support in January 2014\footnote{Online: \url{http://www.allaboutsymbian.com/news/item/18502\_New\_Symbian\_and\_Meego\_applicat.php/} [accessed: January 13\textsuperscript{th}, 2021]}.

\begin{figure}
    \centering
    \includegraphics[width=\textwidth]{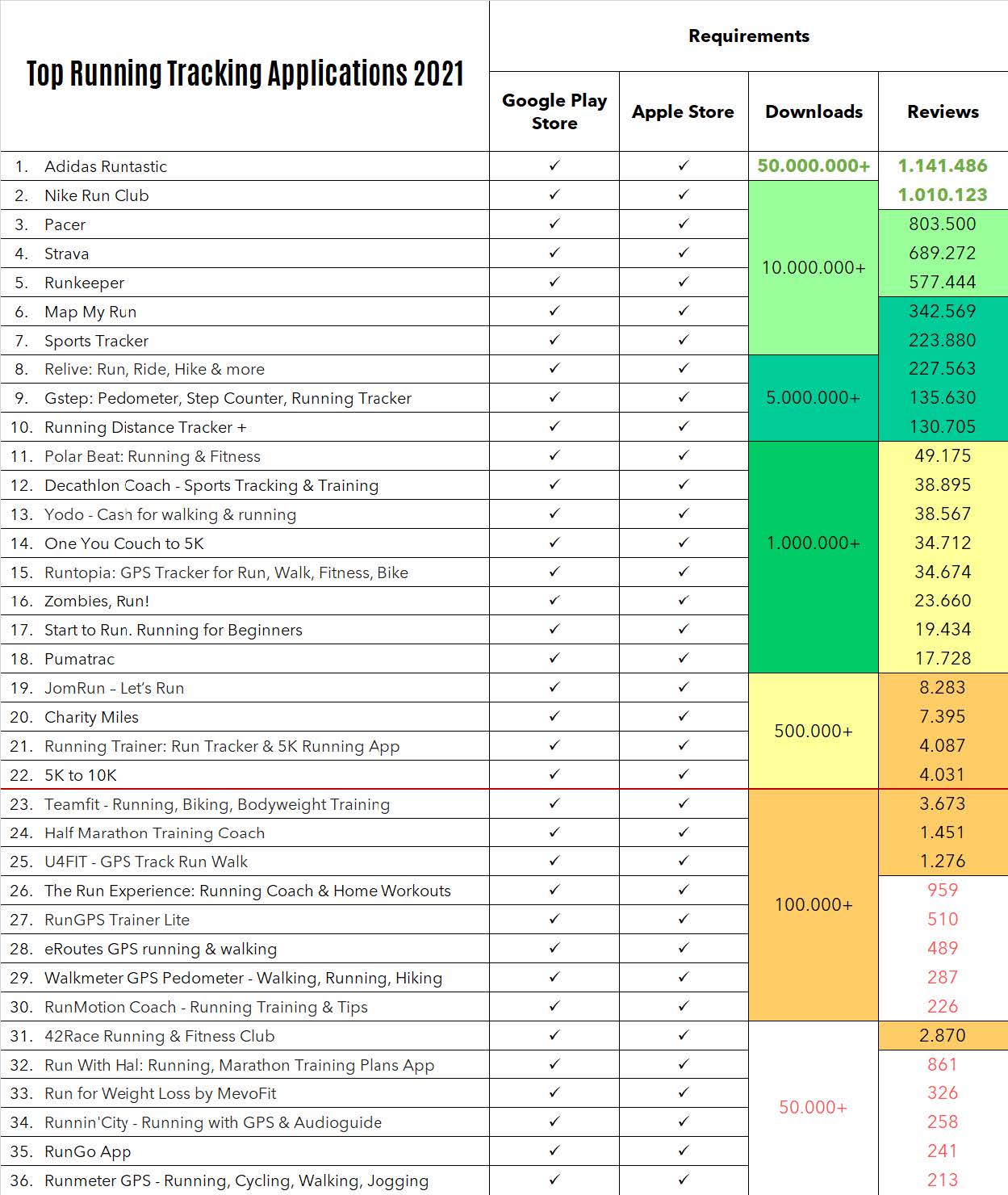}
    \caption{Top Running Tracking Applications in 2021}
    \label{fig:appselection}
\end{figure}

Furthermore, we observed that Apple's iOS store does not disclose the number of downloads or reviews per application, but rather the number of zero-to-five-star ratings and their average. Since research has shown that online zero-to-five-star ratings tend to be predominantly positive and additionally rather unreliable in terms of indicating the success of a product~\cite{Langhe2016NavigatingBT}, we focused solely on Google's play store to check for an app's download and review requirements, presuming that the most popular applications would be available and similarly well-known (i.e., downloaded and reviewed) in Apple's iOS store. Thus, eventually we considered apps that had at least \emph{50K+} downloads and \emph{100+} reviews on Google's Play store, and were also available on iOS. As this resulted in a selection of 36 running tracking apps we increased the number of minimum downloads to \emph{500K+} so that we ended up with 22 running tracking apps for our analysis. Fig.~\ref{fig:appselection} shows all apps which were considered in the first round and where we defined the cut-off.  
All of the 22 selected running tracking applications were installed on an iPhone SE. User profiles were created and after some test runs we started with our feature analysis.


\subsection{Feature Analysis}
In order to investigate the contribution of different application features to intrinsic motivation, we focused on a target population of recreational runners who show little intrinsic motivation. We used the Intrinsic Motivation Inventory (IMI)\cite{Monteiro2015}, which is grounded in Self Determination Theory \cite{ryan1982control} as a screening instrument.
The inventory consists of seven Likert-type subscales, considering one's \emph{interest/enjoyment}, \emph{perceived choice}, \emph{effort}, \emph{value/usefulness}, \emph{pressure/tension},\emph{perceived choice} and \emph{relatedness}, last of which is used as a measure of relevance for interpersonal interactions. 

We were able to recruit $n=15$ participants fitting this sample frame (9 female, AVG age = 24) with whom we conducted semi-structured, problem-centered interviews, using an adaptation of Gutman's laddering technique~\cite{gutman1982means}. According to Schultze \& Avital such a laddering technique \textit{encourages the interviewee to elaborate on the meaning of his/her personal constructs by narratively forging links between them [...] which is achieved by asking how and why questions''}~\cite[p.9]{schultze2011designing}.

Interviewees were asked to talk about different app functions and the consequences they would attach to using them. This technique was repeated until interviewees arrived at a point where they had clearly explained their feelings towards a certain functionality of a feature and its use. Additionally, interviewees were asked to suggest features they believed running tracking applications could integrate to motivate them.

All interviews were recorded, transcribed and subsequently analyzed using Kuckartz's~\cite{kuckartz2014qualitative} structuring content analysis. Main categories were based on the features linked to Bauer and Krieglstein’s framework. Sub-codes were created based on interviewees' perceptions regarding the feature's motivational impact. A second round of analysis assured coding reliability.

\section{Discussion of Results}\label{sec:results}
\begin{figure}
    \centering
    \includegraphics[width=\textwidth]{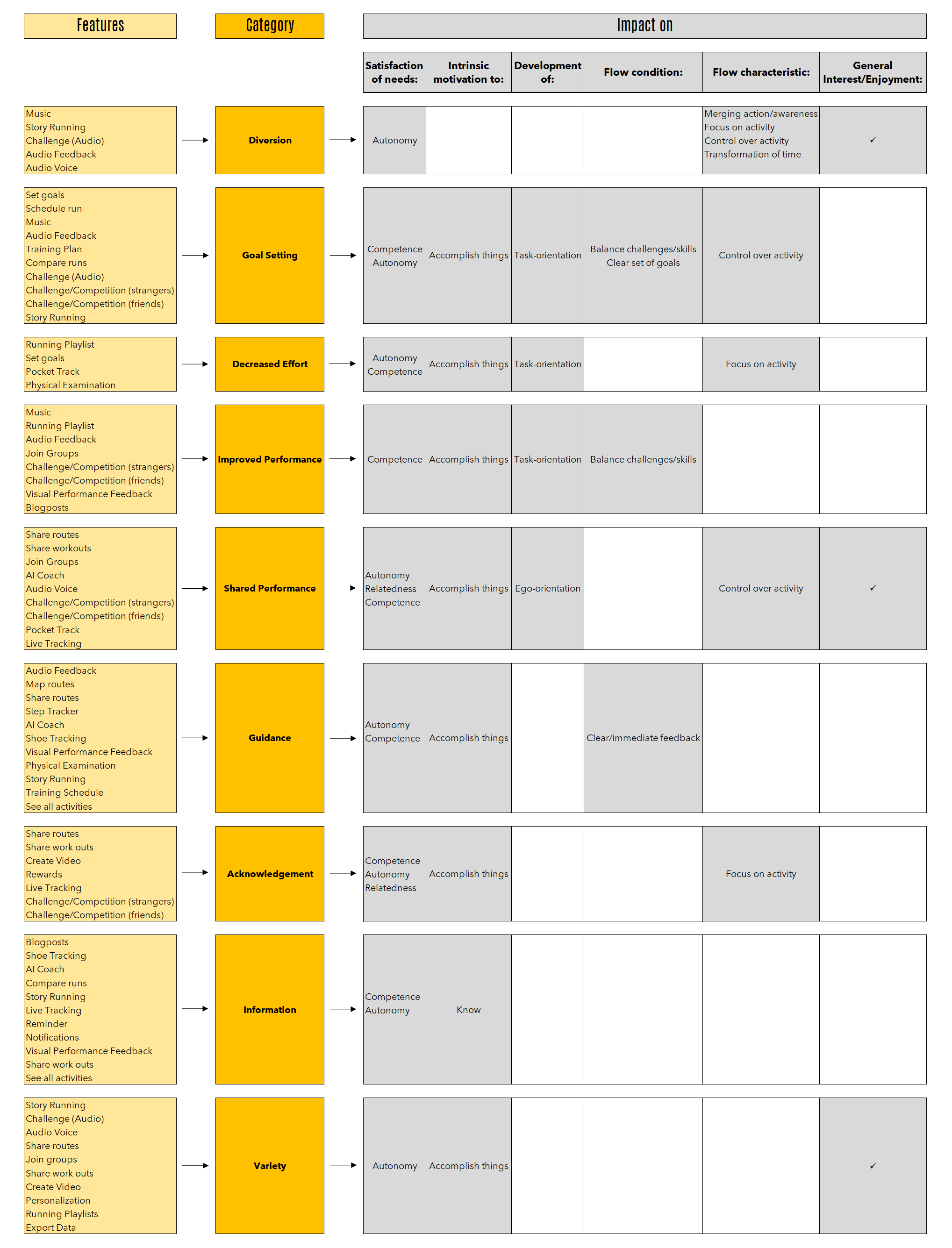}
    \caption{Categorisation of features and their impact}
    \label{fig:results}
\end{figure}

Results show that running tracking application features do contribute to recreational runners’ intrinsic motivation by (1) diverting them, (2) aiding their goal setting, (3) decreasing their efforts, (4) improving and (5) sharing their performance, (6) giving them acknowledgement as well as providing them with (7) guidance, (8) information, and (9) a greater variety in training options. Fig.~\ref{fig:results} shows that, depending on runners' perceptions, these features help satisfy needs, they impact on motivation, they help achieve flow-like experiences, or generally trigger interest or enjoyment. 


For example, runners' intrinsic motivation seems to benefit from features which increase their felt autonomy and perceived competence, such as those which help to individually schedule (and thus not postponed) runs, or those which log individual achievements and thus provide information on runners' change in competence. Features that particularly aid in improving peoples' running performance (e.g., audio feedback by the app, listening to a running playlist that fits their pace, etc.) further boost this perceived competence improvement.

Furthermore, being acknowledged by others (i.e., virtually being cheered on) for achievements or acknowledging oneself, by sharing a run, helps create a certain level of relatedness and consequently encourages runners to keep working on their set goals (cf.~\cite{deci2013intrinsic}). To this end, interviewees also found that sharing one's performance and comparing oneself to others can have significant positive effects on their intrinsic motivation (cf.~\cite{Tauer2004}). Competing may even initiate a certain ego-orientation, which according to Elbe \& Schüler encourages athletes~\cite{elbe2020motivation}. Loosing such a comparison, however, can easily diminish the feeling about one's competence and thus hamper motivation~\cite{duda2001achievement}. 



Interestingly, the setting of external goals, e.g., the completion of a challenge or public competition set to a specific distance, date or time frame, was often perceived as controlling and thus interfering with runners' desire for autonomy. So were features which would provide continuous (i.e., too much) guidance and feedback. 

Also, a permanent push towards exceeding one's running abilities was criticized, as one interviewee puts it: \textit{``Maybe I do not want to run any faster at that moment but have actually just found a great pace where I do not get a stitch or anything else and have a good rhythm''}. 

Some features, such as clearly set goals or (motivating) running playlists, were found to  decrease runners' efforts and thus to increase the likelihood for them to convert their intention into a concrete action, i.e., start their run. Hence, being guided through, e.g., hearing audio feedback, hearing a story or being able to map routes was considered empowering and helpful in preparing runs: \textit{``[W]hen I see the route and it is eight kilometers, for example. Well, eight kilometers is not really eight kilometers. There is altitude in it, there are traffic lights [...] For example, you could make five kilometers in the same duration as the eight kilometers. Then I could know how to best combine duration and distance"}. 

Interviewees also found that features which would foster task-orientation (cf.~\cite{elbe2020motivation}), such as the use of a training plan, would encourage them to take action, or at least think about it, which might also increase one’s intrinsic motivation to accomplish things (cf.~\cite{vallerand1997toward}). 

The opposite effect was described by interviewees when using the pocket track feature, since here they want a conscious decision and therefore to keep autonomy over the beginning of a training. As one interviewee states, pressing the start button and seeing the countdown is like crossing a psychological barrier: \textit{``Yes, it is important because then I know: It [the run] starts now''}. The physical examination feature, however, was described as effort increasing, as interviewees felt it would delay their run and consequently hamper their intention.

In terms of features providing visual guidance and feedback after the run, interviewees agreed on their empowering effect, allowing them to see and potentially disseminate their achievements. Although, some participants would like to hide certain performance values, such as pace or distance, as they feel it would pressure them into performing better.

Interviewees further found that through many of the offered features (e.g., receiving information in the form of blog posts in an application or general visual feedback of their performance) they would learn something new about themselves and their bodies, which increases the intrinsic motivation to know about and better understand personal health (cf.~\cite{vallerand1997toward}); even more so, when they are able to apply this understanding to future runs and thus see an increase in their individual competences. 

We further found that certain features (e.g., features focusing on individual progress rather than external praise) may change the type of motivation, turning extrinsically motivated runners into more intrinsically motivated ones. Other features (e.g., rewards) trigger the opposite, since interviewees see them as tangible acknowledgements of their achievements. To this end, the possibility of donating through their runs has been perceived as particularly valuable by some interviewees, as it would allow them to easily help others. As such a reward is usually not depending on a specific goal, it is also not perceived as controlling and therefore does not hamper one's intrinsic motivation (cf.\cite{ryan2000intrinsic}). 

A variety of features, however, seem to polarize, e.g., story running, creating a video of a run, hearing an actual challenge, or setting the type of voice for the audio feedback. 
While these features were described as interesting, encouraging, enjoyable and helpful in providing training variety, by some participants, others identified them as unauthentic, unnecessary or annoying. And even if they would be curious about them they would soon lose interest.
For example, regarding the type of audio feedback voice one interviewee stated: \textit{``it always seems a bit fake [...] [y]ou can already hear the laughter in their voice although they are not laughing''}. To this end, it was also pointed out that certain voice types may even be disrupting.
Others, however, stated that the audio feedback would motivate and help them focus on the running activity. This direct influence on one's motivation may further trigger the transfer from a situational to a contextual level of intrinsic motivation, where runners' episodic running experiences can change their intrinsic motivation towards running in general~\cite{vallerand1997toward}.

Finally, we found that those features which allow for (1) an appropriate balance between perceived challenges and existing skills (e.g., use of a training plan, or listening to a playlist based on one's pace), (2) a clear goal setting, and (3) immediate feedback, can help runners reach a flow-like experience~\cite{csikszentmihalyi2005handbook}, and therefore may move them towards an optimal state of motivation. This particularly includes features which would help them concentrate on the act of running while at the same time providing them with a sense of control over and awareness for the activity; e.g., features that divert from nearing exhaustion, such as listening to a story or challenge to complete. 



\section{Conclusion and Future Outlook}\label{sec:conclusion}
Summarizing, our investigation has shown that certain features of running tracking applications, e.g.: \emph{using specific running playlists, hearing audio feedback, stories and challenges, setting goals, using training plans, competing against strangers and friends, live tracking or sharing one's workout}, can contribute to recreational runners' intrinsic motivation. Depending on runners' perceptions, these features may have beneficial or detrimental effects. First, they satisfy needs of autonomy, competence and relatedness, which according to Deci \& Ryan~\cite{deci2000and} strongly impact on intrinsic motivation. Second, they let runners know about and accomplish goals, which can help transfer situational and contextual levels of motivation into general motivation to run~\cite{vallerand1997toward}. Third, they trigger task- and ego-orientation, leading to competitiveness which can have both positive (i.e., when winning)~\cite{mcauley1989effects} or negative (i.e., when losing)~\cite{duda2001achievement} effects on intrinsic motivation. Finally, their use may trigger flow-like states through offering control over the activity as well as providing clear and immediate feedback which is in line with peoples' distinct abilities~\cite{schuler2020intrinsische}.

Our analysis focused on recreational runners possessing low levels of intrinsic motivation towards the sport. Hence, results may not generalize to other athlete groups. And also older runners may have very different perceptions concerning the use and potential benefits of running tracking applications used by our analysis (note: the average age of our interviewees was 24). 

Future work should thus aim to close this gap and evaluate the extent to which the above presented insights also apply to other target populations. Furthermore, we recommend expanding upon these findings and focus on features which help create more individual running experiences. Janssen et al., for example, developed an online tool to support users in finding a running tracking application that best fits their individual needs~\cite{janssen2020app}. More such research, which aims to increase the chance of running tracking applications being capable of creating long lasting habits for runners, may eventually ensure that we reach the level of regular physical activity the WHO recommends.

\bibliography{hcii}
\bibliographystyle{splncs04}
\end{document}